\documentclass[letter]{aa}  
\usepackage{graphicx}
\usepackage{txfonts}
\usepackage{gensymb}

\begin{document} 

   \title{Straight outta photosphere: Open solar flux without coronal modeling}
   
   \author{Ismo Tähtinen
          \inst{1}
          \and
          Timo Asikainen\inst{1}
          \and
          Kalevi Mursula\inst{1}
          }

   \institute{Space Physics and Astronomy Research Unit, University of Oulu,
               POB 8000, FI-90014, Oulu, Finland\\
              \email{ismo.tahtinen@oulu.fi}
             }


 
  \abstract
   {The open solar flux, that is, the total magnetic flux escaping the Sun, is one of the most important parameters connecting solar activity to the Earth.
   The open solar flux is commonly estimated from photospheric magnetic field measurements by making model assumptions about the solar corona.
   However, the question in which way the open solar flux is directly related to the distribution of the photospheric magnetic field is still partly unknown.}
   {We aim to reconstruct the open solar flux directly from the photospheric magnetic fields without making any assumptions about the corona and without using coronal hole observations, for instance.}
   {We modified an earlier vector sum method by taking magnetic field polarities into account and applied the method to the synoptic magnetograms of six instruments to determine the open solar flux from solar cycles 21--24.
   }
   {The modified vector sum method produces a vector of the global solar magnetic field whose magnitude closely matches the open solar flux from the potential field source surface (PFSS) model both by the absolute scale and the overall time evolution for each of the six magnetograms. 
   The latitude of this vector follows the Hale cycle by always pointing toward the dominantly positive-polarity hemisphere, and its longitude coincides with the location of the main coronal holes of the McIntosh Archive.
   We find multi-year periods during which the longitude of the vector slowly drifts or stays rather stationary in the Carrington frame.
   These periods are punctuated by times when the longitude moves rapidly in the Carrington frame.
   By comparing the magnitude of this vector to the open solar flux calculated from the PFSS model with different source surface heights, we find that the best match is produced with a source surface height $R_{ss} = 2.4-2.5R_\odot$.}
  {}

\authorrunning{Tähtinen et al.}
   \maketitle
%

\section{Introduction}\label{sec:intro}
The open solar flux (OSF) is the amount of magnetic flux that extends from the Sun into the interplanetary space.
The OSF is determined by the open photospheric magnetic fields that do not return to the solar surface, but are carried away by the solar wind into the interplanetary space, where they form the heliospheric magnetic field (HMF) that spreads throughout the heliosphere.
The HMF plays a crucial role in space weather and space climate applications as it couples the Earth's magnetosphere directly with the Sun, allowing energy transfer from the solar wind into the magnetosphere.
Measuring the HMF is the only way to directly estimate the amount of OSF.
Based on the result that the magnitude of the radial component $B_r$ of the HMF is essentially independent of heliographic latitude \citep{Balogh1995}, the OSF can be directly estimated from the in situ satellite observations as $\Phi = 4\pi R_{AU}^2 |B_r|$.

In addition to the in situ measurements of the HMF, several methods have been developed to estimate the OSF indirectly from solar observations \citep{Wang1995,Solanki2000,Riley2001,Lockwood2014,Lockwood2022}.
Photospheric magnetic fields can be used to estimate the OSF, but a coronal model is needed to extrapolate magnetic fields to the solar corona and to determine which magnetic field lines return to the solar surface and which escape to the interplanetary space.
The simplest such model is the potential field source surface (PFSS) model \citep{Altschuler1969,Schatten1969,Wang1992}, which assumes that the solar atmosphere is current-free and that the field is radial at a certain distance that is called the source surface radius.
In the PFSS model, the OSF is equal to the total unsigned flux at the source surface.
In addition to coronal models, coronal hole observations can also be used to determine the approximate location of open field regions in the photosphere \citep{Lowder2014, Lowder2017, Linker2017}.
So far, the OSF has not been directly related to photospheric observations \citep{Owens2013}.

Since the nineteenth century, it has been known that the solar activity is distributed asymmetrically \citep{Carrington1863}.
The regions of enhanced activity that produce this asymmetry are called, for example, active longitudes, activity nests, or hot points \citep[][]{Warwick1965, Castenmiller1986, Bai1988}.
To quantify the longitudinal asymmetry of the solar activity,
\citet{Vernova2002} used the vector sum method of \citet{Vernov1979} on sunspot observations, and later extended their analysis to solar magnetograms \citep{Vernova2007}.
This method represents the local solar activity on the solar surface as a vector in polar coordinates whose magnitude is determined by the level of activity (e.g., the sunspot area) and the phase angle by the respective Carrington longitude.
The vector sum is the sum of such vectors over all longitudes during one Carrington rotation.
The method is suitable for studying longitudinal asymmetries because uniformly distributed activity averages to zero.

In this Letter, we present a simple and straightforward modification of the vector sum method that produces a vector whose magnitude closely matches the OSF obtained from the PFSS model and directly relates the OSF to photospheric magnetic fields.
We use synoptic observations of the photospheric magnetic fields from six observatories for solar cycles 21--24 and analyze them with the modified vector sum method.
This article is structured as follows.
We describe the data in Sec.~\ref{sec:data} and introduce the modified vector sum method in Sec.~\ref{sec:method}.
In Sec.~\ref{sec:VectorSumFlux} we compare the vector sum flux to PFSS OSF from six instruments and relate the evolution of the vector to the Hale cycle.
In Sec.~\ref{sec:long} we analyze the evolution of the vector sum longitude.
In Sec.~\ref{sec:SourceSurface} we calculate the optimum PFSS source surface radius based on the vector sum flux.
We discuss our results and give our conclusions in  Sec.~\ref{sec:Disc}.

\section{Data}\label{sec:data} 
We used six datasets of synoptic maps of the photospheric magnetic field: from the Mount Wilson observatory \citep[MWO;][]{Howard1974}, from Kitt Peak \citep[WSO;][]{Livingston1976,Jones1992}, from the Wilcox solar observatory \citep[WSO;][]{Scherrer1977}, from the Michelson Doppler Imager on board the Solar and Heliospheric Observatory \citep[SOHO/MDI;][]{Scherrer1995}, from the Vector Spectromagnetograph on the Synoptic Optical Long-term Investigations of the Sun telescope \citep[SOLIS/VSM;][]{Keller2003}, and from the Helioseismic and Magnetic Imager on board the Solar Dynamics Observatory \citep[SDO/HMI;][]{Scherrer2012,Pesnell2012} (see \citealp{Virtanen2016} for a detailed review of datasets).
We scaled the of MWO, KP, MDI, SOLIS, and HMI synoptic maps to a common resolution of 180 pixels in sine of latitude and 360 pixels in longitude.
The resolution of WSO data is 30 pixels in sine of latitude and 72 pixels in longitude.
The data cover Carrington rotations 1617--2284 starting in the late declining phase of solar cycle 20 (July 1974) and reaching the current solar maximum (May 2024).
We converted line-of-sight measurements into a pseudo-radial field by dividing with the cosine of latitude and calculated the PFSS OSF using source surface radius $R_{ss} = 2.5R_\odot$ for the comparison with the vector sum flux.
The highest term used in harmonic expansion was $n_{max} = 9$ for the WSO data and $n_{max} = 50$ for other data.

The pixel value in a synoptic map represents the average signed magnetic field strength within the pixel.
We converted the pixel values of the magnetic field into (signed) magnetic fluxes by multiplying them with the corresponding pixel area for the vector sum calculation.

\begin{figure}
\centering
{\includegraphics[height=1\textheight,keepaspectratio]{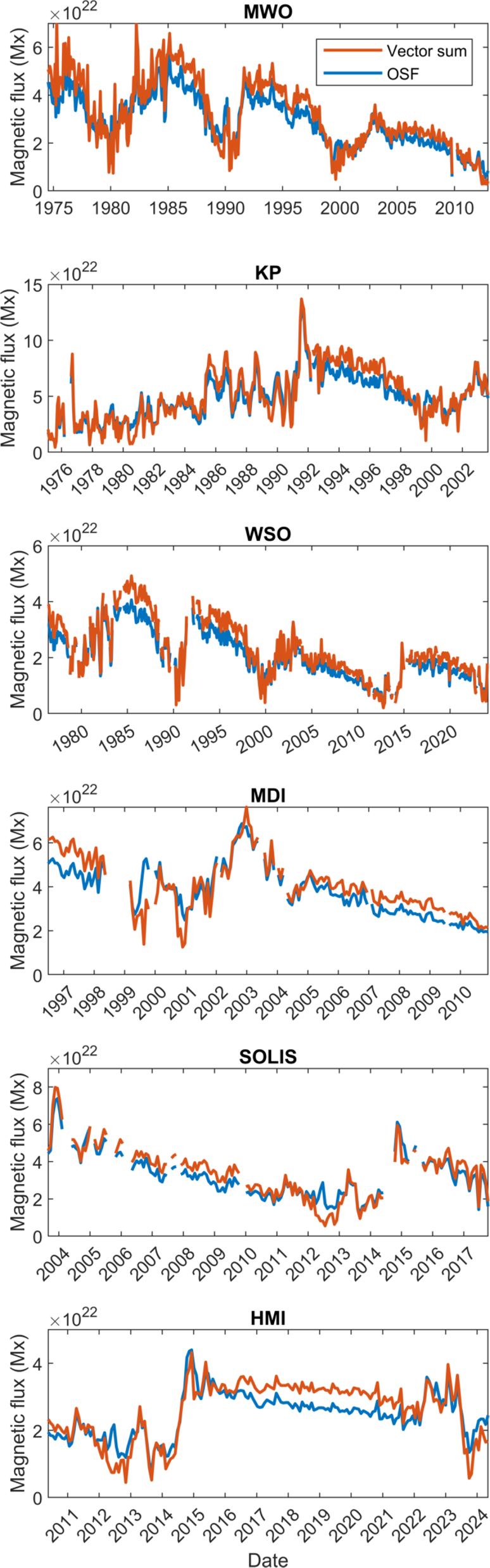}}
 \caption{Vector sum flux (orange) and the PFSS OSF (blue) for six different datasets.}
 \label{fig:magnitdue}%
\end{figure}

\section{Vector sum}\label{sec:method}
The vector sum method we use in this paper was modified from the method used by \citet{Vernova2007} for magnetogram data.
While \citet{Vernova2007} used the unsigned magnetic field strength, we took the polarity of the magnetic field into account.
\citet{Vernova2007} also averaged their data over latitudes and only considered the longitudinal distribution of magnetic fields, which they represented in polar coordinates.
We extended this two-dimensional method to three dimensions by representing all magnetogram data in spherical coordinates.

The vector sum was calculated by first representing each pixel of a magnetogram in spherical coordinates, where the unsigned magnetic flux represents the length of the vector, and the latitude and longitude of the magnetogram correspond to copolar and azimuthal angles in spherical coordinates.
The vector sum of a synoptic magnetogram is defined straightforwardly as the sum of these (signed) vectors.
The vector sum is related to the asymmetric distribution of the unipolar magnetic field in the photosphere.
The magnitude of this vector, called the vector sum flux here, is related to the total asymmetric unipolar flux, and its direction is associated with the mean location of the asymmetric unipolar flux.

Since multiplying a vector with~-1 reverses its direction, negative polarities in a magnetogram correspond to the positive polarities of equal magnitude on the opposite side of the Sun.
Thus, opposite polarities on opposite sides of the Sun interfere constructively.
Large unipolar regions contribute much to the vector sum, and nearby opposite polarities almost cancel out.
For example, the vector sum of a bipolar magnetic region is close to zero.
(The vector sum is also zero when the unipolar magnetic field is uniformly distributed on the surface. Thus, the possible monopole term is automatically canceled out.)

\section{Vector sum flux and Hale cycle}\label{sec:VectorSumFlux}
Figure \ref{fig:magnitdue} shows the vector sum flux (orange) together with the OSF from the PFSS ($R_{ss} = 2.5R_{\odot}$) model (blue) for six different datasets.
The vector sum flux not only closely follows the time evolution of the PFSS OSF, but the magnitude of the vector sum flux also matches the OSF strength for all datasets.
As has been known for a long time, there are large differences between the PFSS OSF of the six datasets, especially during the earlier times. They are due to differences in observational methods, resolutions, and instrumentations \citep[see, e.g.,][and references therein]{Pietarila2013,Riley2014,Virtanen2016}.
However, Fig.~\ref{fig:magnitdue} demonstrates that the vector sum method produces a magnetic flux for each of the six datasets whose strength closely matches the PFSS OSF, despite these differences.
The vector sum flux and the PFSS OSF also differ somewhat.
The largest differences are seen in the declining and minimum phases of the solar cycle.
During these times, the vector sum tends to slightly overestimate the magnetic flux compared to the PFSS.
However, the PFSS OSF may not necessarily be more correct than the vector sum method. Comparisons with the HMF have shown that the PFSS OSF misses some flux at these times.
\cite{Lee2011} noted that during the two studied solar minima, the optimum PFSS source surface radius should be smaller, about 1.8-1.9, leading to a higher flux than for a larger constant radius at all times.
\citet{Virtanen2020} found that the source surface radius varies in time.
Accordingly, the question of whether the vector sum or the PFSS OSF is more correct cannot yet be answered.

\begin{figure}[!hbt]
	\centering
	\resizebox{\hsize}{!}{\includegraphics[width=\textwidth,height=\textheight,keepaspectratio]{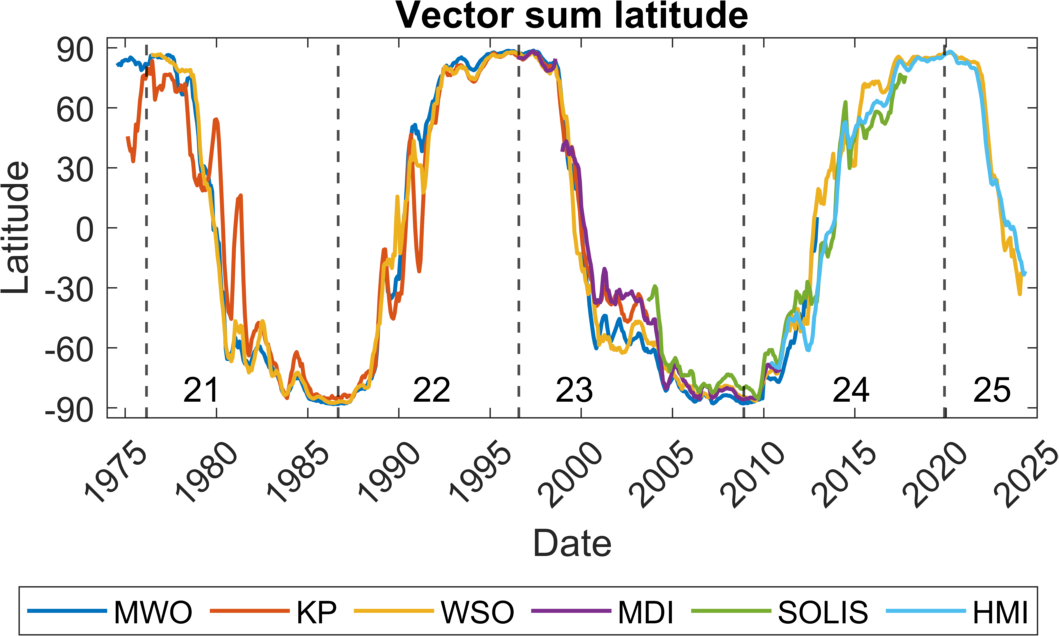}}
	\caption{Vector sum latitude smoothed with a seven-rotation running-mean window. The dashed vertical black lines show the location of solar minima. The numbers denote the solar cycle number.}
	\label{fig:lats}%
\end{figure}

Figure \ref{fig:lats} shows the vector sum latitude calculated for the six datasets.
It is evident that the vector sum latitude follows the Hale cycle.
The start of our dataset corresponds to the time of solar minimum in 20/21 during which the sign of the north pole was positive (positive-polarity minimum).
Accordingly, the vector sum points north, as depicted by the vector sum latitude.
As the solar cycle proceeds toward the solar maximum, the toroidal component of the solar magnetic field becomes dominant, and the vector sum travels toward the equator.
The vector sum continues toward the south pole as the opposite-signed polar fields of the upcoming minimum start to form.
During the minimum between cycles 21 and 22, the vector sum points south, in the direction opposite to the previous minimum.
During the next cycle, the vector sum again turns to the north pole, thus completing one Hale cycle.

\section{Evolution of the vector sum longitude}\label{sec:long}
\begin{figure*}[!hbt]
\centering
 \resizebox{\hsize}{!}{\includegraphics[width=\textwidth,height=\textheight,keepaspectratio]{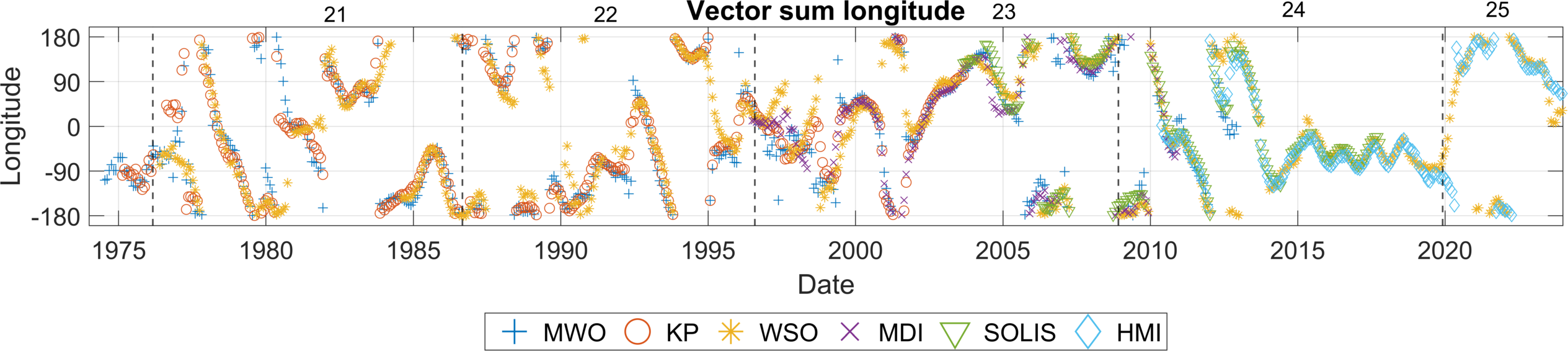}}
 \caption{Vector sum longitude smoothed with a seven-rotation running-mean window. The dashed vertical black lines show the location of solar minima. The numbers in the plot show the solar cycle number.}
 \label{fig:longs}%
\end{figure*}

Figure~\ref{fig:longs} shows the vector sum longitude for the six datasets.
The closely matching data points in Fig.~\ref{fig:longs} show evidence for a detailed agreement in the azimuthal distribution and longitudinal asymmetry between different datasets.
There are long periods during which the vector sum longitude is stationary or drifts consistently with respect to the Carrington frame.
For example, starting from the maximum of cycle 24 in late 2014, the vector sum longitude stayed within one quadrant of the Carrington longitude for the rest of this cycle for about five years.
On the other hand, the vector sum longitude drifted quite systematically with respect to the Carrington longitude in 2001-2004.
The vector sum longitude can drift in the retrograde (slower) and prograde (faster) direction in the Carrington frame, but retrograde motion is more common, with 58\% of longitudinal movements between consecutive rotations being retrograde.
From a linear fit, we find that the drift speed of average vector sum longitude ranges from -41.1\degree{} retrograde (CR2096-CR2100) to 12.5\degree{} prograde (CR1984-CR1990) per Carrington rotation.
These drift speeds correspond to rotation rates of 408 nHz (28.4 days) and 471 nHz (24.6 days).
Stationary and slowly evolving periods are interrupted by more chaotic times during which the longitude makes large changes more abruptly.
These times tend to be around solar maxima, as in 1980, 1989-1991, and 2001, but they can also occur in other phases of the solar cycle, as in 1995 and 2020.

\begin{figure}
\centering
\resizebox{\hsize}{!}{\includegraphics[width=\textwidth,height=\textheight,keepaspectratio]{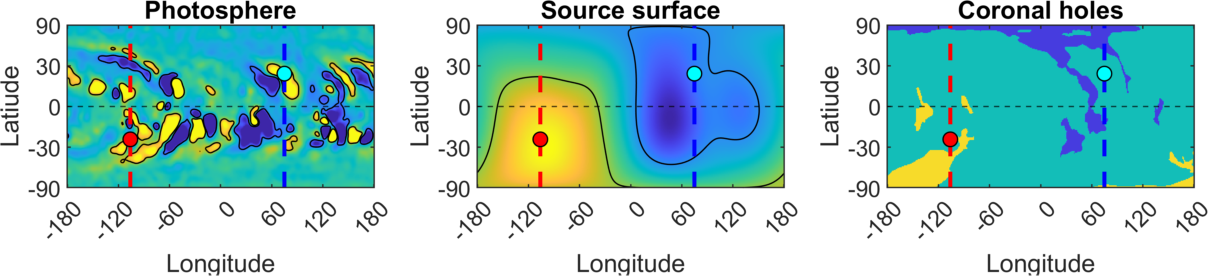}}
\caption{KP synoptic maps for Carrington rotation 1999. Left column: Photospheric magnetic field. The magnetogram was smoothed with a Gaussian smoothing kernel with a standard deviation of 3. The black lines correspond to $B = \pm10$G contours. Center column: KP PFSS source surface magnetic field. The black lines correspond to $B = \pm0.15$G contours. Right column: Coronal holes from the McIntosh archive. Yellow and blue correspond to positive and negative polarity in all maps. The dashed red and blue lines mark the vector sum longitude and its opposite. The red and cyan circles mark the vector sum direction and its antipode.}
\label{fig:maps}%
\end{figure}

Figure \ref{fig:maps} shows the KP photospheric magnetic field (left panel), KP PFSS source surface magnetic field (middle panel), and coronal holes (right panel) for Carrington rotation 1999 (February 2003).
The middle panel of Fig. \ref{fig:maps} depicts the PFSS  coronal magnetic field at the $R_{ss} = 2.5$ source surface. It reveals that the vector sum longitude corresponds to the dominant positive magnetic fields at the source surface, while its inverse corresponds to dominant negative magnetic fields.
Likewise, the right panel of Fig.~\ref{fig:maps} of the coronal holes from the McIntosh Archive \citep{McIntosh1964} shows that the largest coronal hole extensions appear close to the vector sum longitude.
We note that it would be difficult to see from the complex distribution of photospheric strengths (left panel) where the coronal unipolar longitudes (middle panel) or coronal hole extensions (right panel) are located.
The success of the vector sum to reproduce these longitudes shows the detailed agreement with the coronal magnetic structure of the PFSS model.

\section{Optimum PFSS source surface radius}\label{sec:SourceSurface}
\begin{figure}
\centering
 \resizebox{\hsize}{!}{\includegraphics[width=\textwidth,height=\textheight,keepaspectratio]{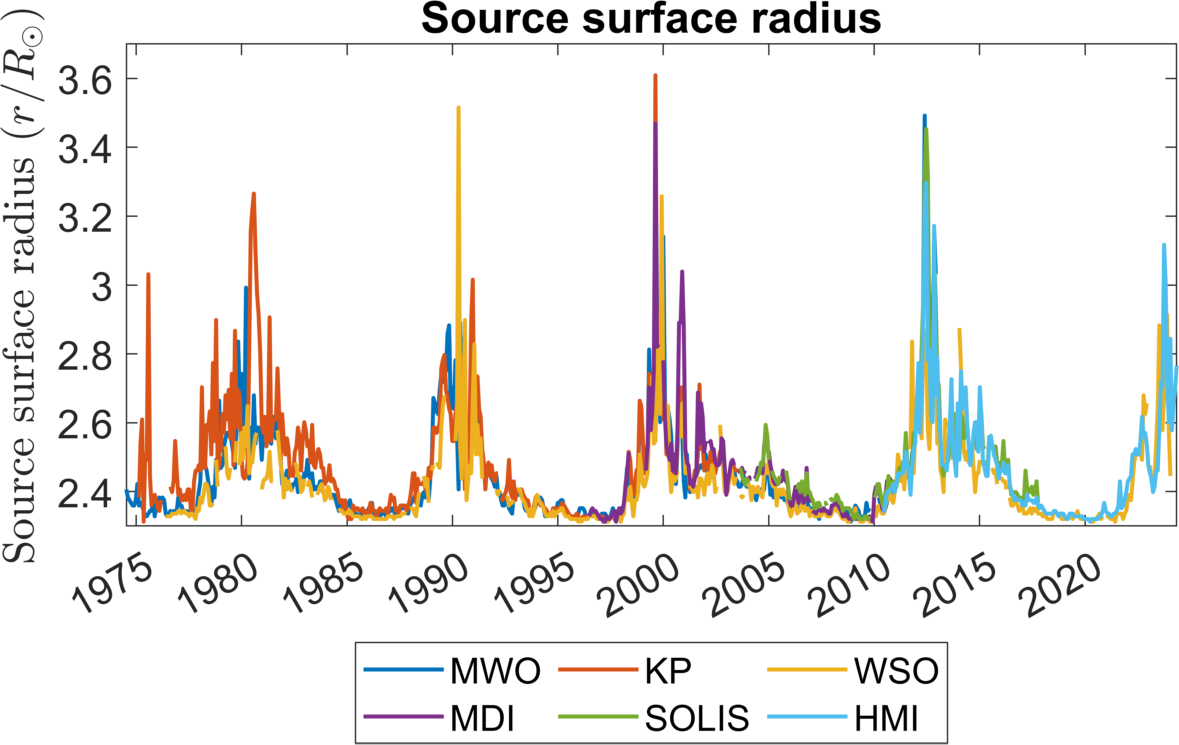}}
 \caption{Optimum PFSS source surface radii for the six datasets.}
 \label{fig:sourcesurface}%
\end{figure}

Since the vector sum provides an independent and parameter-free estimate of the OSF, it can be used to constrain the only free parameter of the PFSS model, the source surface radius.
We calculated for each of the six datasets the source surface radius that produced the closest match with the vector sum flux and the respective PFSS OSF for each rotation.
Figure~\ref{fig:sourcesurface} shows these rotational optimum source surface radii for all datasets.

The smallest rotational optimum source surface radius is 2.31$R_\odot$ for all datasets, except for SOLIS, for which it is 2.32$R_\odot$.
The largest optimum solar source surface radius ranges from 3.30$R_\odot$ (HMI) to 3.61$R_\odot$ (KP).
The mean optimum source surface radii calculated for each dataset over their respective full intervals range between 2.42$R_\odot$ (WSO) and 2.48$R_\odot$ (KP, SOLIS, and HMI). Despite the very different temporal extent and time intervals, the optimum radii for the six datasets agree very well with each other.
In agreement with the discussion in Sec. \ref{sec:VectorSumFlux} (see also \citealp[]{Lee2011} and \citealp[]{Virtanen2020}), the optimum source surface radii were found to change with the solar cycle. They are smallest during solar minima and largest at solar maxima.

\section{Discussion}\label{sec:Disc}
The modified vector sum method produces a vector whose magnitude closely matches the OSF calculated from the PFSS model, both according to its absolute scale and to the time evolution without any additional scaling or adjustments.
This vector sum closely follows the Hale cycle movement of the large-scale dipole field (Fig.~\ref{fig:lats}).

The vector sum method provides a consistent way to calculate the orientation of the large-scale dipole and the corresponding longitude.
We showed that the vector sum direction corresponds to the location of large unipolar regions and is related to the location of coronal holes (Fig.~\ref{fig:maps}).
We found a few multi-year periods during which the vector sum longitude was stationary or evolved consistently in the Carrington frame.
These persistent longitudinal structures have previously been observed by \citet{Carrington1863} and were since then been observed in several solar and heliospheric parameters, including sunspots \citep{Castenmiller1986,Harvey1993,Berdyugina2003,Usoskin2005,Zhang2011b,Gyenge2014}, solar flares \citep{Warwick1965,Bai1987, Bai1988, Bai2003, Zhang2011a, Zhang2015, Gyenge2016}, coronal mass ejections \citep{Gyenge2017}, and in the heliospheric magnetic field \citep{Takalo2002}.
Recently, persistent longitudinal activity seen in coronal bright points and low-latitude coronal holes has been related to magnetohydrodynamic Rossby waves that can propagate in either retrograde or prograde directions \citep{McIntosh2017,Krista2018,Harris2022}.
The observed minimum rotation rate of 408 nHz (28.4 days) corresponds to retrograde motion below about 50\degree{} latitude, and the rotation rate of 471 nHz (24.6 days) corresponds to the rotation rate near the equator ($\theta$ < 10\degree{}) and to prograde rotation at higher latitudes \citep{Snodgrass1990}.
The results presented here based on the vector sum method support the idea that the evolution of large-scale field as observed on the photosphere is possibly linked with the dynamics of magnetohydrodynamic Rossby waves in the tachocline.

By comparing the vector sum flux to the PFSS OSF, we found that the best match between the two is obtained when the source surface radius is set to 2.4$R_\odot$-2.5$R_\odot$.
These results support using the typical value of $R_{ss}$ = 2.5$R_\odot$ in the PFSS modeling.

We conclude that this study is the first to show that the open solar flux can be directly related to photospheric magnetic fields without any assumptions about the solar corona and without any coronal modeling.
The connection found here between the vector sum longitude of photospheric magnetic fields and the open solar flux indicates a new way of understanding the long-term solar variability and estimating solar-terrestrial effects.

\begin{acknowledgements}
I.T. and T.A. acknowledge the ﬁnancial support by the Research Council of Finland to the SOLEMIP (project no. 357249).
I.T. acknowledges the ﬁnancial support by the Finnish Academy of Science and Letters (Väisälä Fund) and the Jenny and Antti Wihuri Foundation.
This study includes data from the synoptic program at the 150-foot Solar Tower of the Mt. Wilson Observatory.
The Mt. Wilson 150-Foot Solar Tower is operated by UCLA, with funding from NASA, ONR and NSF, under agreement with the Mt. Wilson Institute.
Wilcox Solar Observatory data used in this study were obtained via the web site \url{http://wso.stanford.edu} courtesy of J.T. Hoeksema. 
NSO/Kitt Peak magnetic data used here are produced cooperatively by NSF/NOAO, NASA/GSFC and NOAA/SEL.
Data were acquired by SOLIS instruments operated by NISP/NSO/AURA/NSF. SOHO/MDI is a project of international cooperation between ESA and NASA.
HMI data are courtesy of the Joint Science Operations Center (JSOC) Science Data Processing team at Stanford University.
\end{acknowledgements}

\bibliography{bibliography} 

\end{document}